\documentstyle[preprint,aps]{revtex}

\begin{document}

\draft

\title{Quantum effects in the Alcubierre warp drive spacetime}

\author{William A.\ Hiscock\footnote{e-mail: hiscock@montana.edu}}

\address{Department of Physics, Montana State University, Bozeman,
Montana 59717-3840}

\date{July 9, 1997}
\preprint{MSUPHY97.12}
\maketitle

\begin{abstract}

The expectation value of the stress-energy tensor 
of a free conformally invariant scalar field is computed in
a two-dimensional reduction of the Alcubierre ``warp drive'' spacetime.
The stress-energy is found to diverge if the apparent velocity of the
spaceship exceeds the speed of light. If such behavior occurs in
four dimensions, then it appears implausible that ``warp drive''
behavior in a spacetime could be engineered, even by an arbitrarily
advanced civilization.

\end{abstract}
\newpage

Alcubierre\cite{alc} has described a spacetime which has several of
the properties associated with the ``warp drive'' of science fiction. By
causing the spacetime to contract in front of a spaceship, and expand
behind, the Alcubierre warp drive spacetime allows a spaceship to
have an apparent speed relative to distant objects much greater than
the speed of light.

The stress-energy needed to have a spacetime of this sort is known to
require matter which violates the weak, strong, and dominant energy
conditions\cite{alc}. While quantized fields
can locally violate the energy conditions, Pfenning and Ford\cite{FS}
have recently demonstrated that the configuration of exotic matter 
needed to generate the warp ``bubble'' around the spaceship is quite 
implausible. 

In this letter, a different issue involving quantum effects and the
warp drive spacetime is examined. The curved spacetime associated
with the warp drive will create a nonzero expectation value
for the stress-energy of a quantized field in that spacetime. This field
is assumed to be a spectator in the spacetime, not responsible for
the stress-energy which supports the exotic warp drive metric. While
calculating the expectation value of the stress-energy of a quantized
field in a spacetime is generally an extremely difficult task, the
work involved is greatly reduced if one confines attention to a
two-dimensional spacetime. The warp drive spacetime admits a natural
two-dimensional reduction containing the worldline of the spaceship.
A coordinate transformation then renders the
two-dimensional metric into a static form. For a conformally
invariant massless quantized scalar field, the stress-energy is
then completely determined by the trace anomaly, conservation,
and the values of two integration constants which are determined
by the state of the field\cite{DFU,LHA}.

The resulting expressions for $\langle {T_\mu}^\nu\rangle$
are found to be everywhere regular so
long as the ship does not exceed the speed of light, $v < 1$.
However, for apparent ship velocities exceeding the speed of
light, the stress-energy diverges at a particular distance
from the ship dependent upon the speed. This divergence is
associated with an event horizon which forms in the two-dimensional
spacetime. If the instability is not an artifact of working in two
dimensions, then the spaceship would presumably be precluded
from attaining apparent velocities greater than light due to
metric backreaction effects. 

The warp drive metric proposed by Alcubierre may be written as
\begin{equation}
ds^2=-dt^2+(dx-vf(r)dt)^2+dy^2+dz^2  ,
\label{Ametric}
\end{equation}
where $v$ is the apparent velocity of the spaceship,
\begin{equation}
v={dx_s(t) \over dt}  ,
\label{vdef}
\end{equation}
$x_s(t)$ is the trajectory of the spaceship (chosen to be along
the $x$ direction), $r$ is defined by
\begin{equation}
r=\left[(x-x_s(t))^2+y^2+z^2 \right]^{1/2}  ,
\label{rdef}
\end{equation}
and $f$ is an arbitrary function which decreases from unity at
$r=0$ (the location of the spaceship) to zero at infinity. 
Alcubierre gave a particular example of such a function,
\begin{equation}
f_A (r) = {{\tanh(\sigma(r+R))-\tanh(\sigma(r-R))} \over
        {2 \tanh(\sigma R)}},
\label{Alf}
\end{equation}
where $\sigma$ and $R$ are positive arbitrary constants.

In this letter, the function $f$ will not be constrained to
the particular choice made by Alcubierre; $f$ may be chosen
arbitrarily subject only to the boundary conditions at $r = 0$
and infinity. In order to simplify the analysis of the effects
of the spacetime on the quantized field, the velocity of the
spaceship will be taken to be constant, $v = v_0$, which then
implies that
\begin{equation}
        x_s(t) = v_0 t  ,
\label{xst}
\end{equation}
and hence
\begin{equation}
        r= \left[(x-v_0 t)^2+y^2+z^2 \right]^{1/2} .
\label{rgin}
\end{equation}

While the warp drive spacetime is not spherically symmetric, there is
an obvious way to reduce the spacetime to two dimensions. The spacetime
is cylindrically symmetric about the axis $y=z=0$. The two-dimensional
spacetime which includes the symmetry axis also contains the entire
world line of the spaceship. The two-dimensional metric is then
\begin{equation}
        ds^2=-(1-{v_0}^2 f^2)dt^2-2 v_0 f dt dx +dx^2  .
\label{g2d}
\end{equation}
After setting $y=z=0$, $r$ reduces to
\begin{equation}
        r = \sqrt{(x-v_0 t)^2}  .
\label{r2d}
\end{equation}
If attention is restricted to the half of the
spacetime to the past of the spaceship ($x > v_0 t$), then the
square root in Eq.(\ref{r2d}) may be taken, so that in this
domain, $r = x-v_0 t$ (results for the other half-space may be
obtained by a trivial transformation).

Since the spaceship is traveling with constant velocity, there should
exist a Lorentz-like transformation to a frame in which the ship is
at rest.  The required transformation is most easily
understood if broken into several steps. First, since the metric
components only depend on the quantity $r$, it is natural and
possible to adopt this as a coordinate, transforming from $(t,x)$
coordinates to $(t,r)$ coordinates by making the replacement
$ dx = dr +v_0 dt$ in the metric of Eq.(\ref{g2d}). This yields
\begin{equation}
        ds^2 = -A(r)\left(dt - {{v_0 (1-f(r))} \over A(r)}dr
        \right)^2+{dr^2 \over A(r)}  ,
\label{g2}
\end{equation}
where
\begin{equation}
        A(r) = 1-{v_0}^2(1-f(r))^2  .
\label{Adef}
\end{equation}
Next, the metric is brought into a comoving, diagonal form by
defining a new time coordinate,
\begin{equation}
        d\tau  = dt - {{v_0 (1-f(r))} \over A(r)}dr ,
\label{taudef}
\end{equation}
which gives the metric form
\begin{equation}
        ds^2 = - A(r) d\tau^2 +{1 \over A(r)} dr^2  .
\label{g3}
\end{equation}
This form of the metric is manifestly static. The coordinates
have an obvious interpretation in terms of the occupants
of the spaceship, as $\tau$ is the ship's proper time
(since $A(r) \rightarrow 1$ as $r \rightarrow 0$). On the
other hand, the coordinates are not asymptotically normalized
in the usual fashion; for large $r$, far from the spaceship,
$A(r)$ approaches $1-{v_0}^2$ rather
than unity. This may be corrected by defining yet one
more set of coordinates, $(T,Y)$, such that
\begin{equation}
        T = \sqrt{1-{v_0}^2}\; \tau   , \quad
        Y = {r \over \sqrt{1-{v_0}^2}} .
\label{TYdef}
\end{equation}
The combined coordinate transformations taking $(t,x)$ into
$(T,Y)$ have the asymptotic form of a Lorentz transformation far from
the spaceship, at large $r$ (or, equivalently, $Y$). In this limit,
\begin{equation}
        T = \gamma(t - v_0 x)  ,\quad  Y = \gamma(x - v_0 t) ,
\label{LT}
\end{equation}
where $\gamma$ is the usual special relativistic factor, $\gamma =
1/\sqrt{1-{v_0}^2}$. The transformations to
$T$ and $Y$ will include a factor $i$ when $v_0 >1$. This is
an obvious consequence of transforming to the comoving frame when
the apparent velocity exceeds unity. While there are no real
complications associated with this transformation, the worry of
even possibly having to deal with complex quantities will be
avoided by using the $(\tau,r)$ coordinate system rather than
the $(T,Y)$ system.

Examining the form of the metric of Eq.(\ref{g2}), the coordinate
system is seen to be valid for all $r > 0$ if $v_0 < 1$. If
$v_0 >1$, then there is a coordinate singularity (and event horizon)
at the location $r_0$ such that $A(r_0) = 0$, or,
\begin{equation}
        f(r_0) = 1 - {1 \over v_0}  .
\label{fhor}
\end{equation}
In this case ($v_0 >1$), the spacetime is somewhat like DeSitter space.
There exists an event horizon such that the static region of the 
spacetime is inside the horizon ($r < r_0$), and the horizon first 
appears at infinity and moves inward as the metric's adjustable 
parameter ($v_0$ or the cosmological constant, $\Lambda$) is increased.

The determination of the stress-energy tensor for a quantized
conformally invariant scalar field in the spacetime of Eq.(\ref{g2})
is now straightforward\cite{LHA}. Integration of the conservation
equation and knowledge of the trace anomaly quickly gives
\begin{equation}
        {T_\tau}^r = C_1   ,
\label{Ttr}
\end{equation}
\begin{equation}
        {T_\alpha}^\alpha = - {A'' \over {24\pi}}  ,
\label{trace}
\end{equation}
\begin{equation}
        {T_r}^r = {{C_2 +[A'(r_0)]^2} \over {96\pi A(r)}}-
        {(A')^2 \over{96\pi A(r)}}  ,
\label{Trr}
\end{equation}
where a prime denotes differentiation with respect to $r$ and
expectation value brackets have been suppressed for notational
simplicity. The
remaining components are trivially related to those given above,
${T_\tau}^\tau = {T_\alpha}^\alpha - {T_r}^r$, and ${T_r}^\tau
= -C_1/A^2$. The integration constants $C_1$, $C_2$, and $A'(r_0)$
are determined by the choice of quantum state for the field.

If the field is assumed to be in a time independent and asymptotically
empty state (the usual Minkowski vacuum state) at large $r$, so that
\begin{equation}
        \lim_{r \rightarrow \infty} \langle {T_\mu}^\nu \rangle
        =0  ,
\label{Tmnasym}
\end{equation}
then, since $A(r) \rightarrow 1-{v_0}^2$ and $A'(r) \rightarrow 0$
as $r \rightarrow \infty$, this requires that
\begin{equation}
        C_1 = C_2+[A'(r_0)]^2 = 0  .
\label{c1c2val}
\end{equation}
With this choice of state, only the diagonal components of the
stress-energy are nonzero. They take on the simple forms:
\begin{equation}
        {T_r}^r = -{(A')^2 \over{96\pi A(r)}}  ,
\label{Trr0}
\end{equation}
\begin{equation}
        {T_\tau}^\tau = - {A'' \over {24\pi}}
        +{(A')^2 \over{96\pi A(r)}}  .
\label{Ttt0}
\end{equation}

If $v_0 < 1$, then the function $A(r)$ is everywhere bounded and
positive, and hence the $(\tau,r)$ coordinate system is regular.
Examination of ${T_\mu}^\nu$ as given in Eqs.(\ref{Trr0},\ref{Ttt0})
shows that the components are everywhere finite.

If $v_0 >1$, then there is an event horizon in the spacetime where
$A(r_0) = 0$; the $(\tau,r)$ coordinate system suffers a coordinate
singularity there. In order to determine the regularity of $\langle
{T_\mu}^\nu\rangle$, it is necessary to evaluate the components in
a frame regular at the horizon. There are several different ways
this may be accomplished. The original $(t,x)$ coordinate system
is regular across the horizon. Unfortunately, however, the expressions
for the components of $\langle{T_\mu}^\nu\rangle$ are 
long, complicated, and not particularly illuminating in this 
coordinate system. Alternately, one may evaluate the
stress-energy components in an orthonormal frame attached to a freely
falling observer. The procedure described in Ref. \cite{LHA} may be
followed to set up such a frame is in the static
metric of Eq.(\ref{g2}). Near the horizon, the observed energy density
will be proportional to
\begin{equation}
        \langle\rho\rangle \sim {{T_r}^r - {T_\tau}^\tau \over A(r)} =
        {-A'' \over {24 \pi A}}-{(A')^2 \over {48 \pi A^2}} .
\label{key}
\end{equation}
Expanding Eq.(\ref{key}) near the horizon, and expressing the
result in terms of the original function $f$, yields
\begin{equation}
        \langle\rho\rangle \sim {-(f')^2 \over {48 \pi}}
        \left[ f - \left(1-{1 \over v_0}\right)\right]^{-2}
        + ... \quad ,
\label{key2}
\end{equation}
where the ellipsis denotes less divergent terms. Clearly, there
is no choice of function $f$ which will cause the leading term
in Eq.(\ref{key2}) to be finite as $f \rightarrow 1-1/v_0$. 

This divergence has a simple origin. The event horizon which forms
when the ship's velocity exceeds unity has a natural temperature
associated with it,
\begin{equation}
	T_{Hawking} = {\kappa \over {2\pi}} = {A'(r_0) \over {4\pi}}
	= v_0{f'(r_0) \over {2\pi}}  .
\label{Temp}
\end{equation}
If the quantum state is chosen to be asymptotically empty (essentially
the Boulware vacuum state), then the temperature of the surrounding
universe does not match the natural temperature of the black hole.
It is then inevitable that the stress-energy of a quantized field will 
diverge on the horizon.

Since the ``warp drive'' spacetime is assumed to be associated with
an intelligent engineering effort rather than an astrophysical cause,
it is legitimate to ask whether the divergence might be ``engineered''
away. Presumably the spaceship designers and engineers could, for
example, control
the shape of the function $f(r)$. However, examination of
Eq.(\ref{key2}) shows that the stress-energy of the quantized field
will diverge on the event horizon regardless of the form of $f$.
In a self-consistent solution of the semiclassical Einstein
equations, the backreaction to this divergence would presumably
prevent the spaceship from achieving an apparent velocity exceeding
the speed of light.

The divergence on the horizon occurs because natural quantum state of
the field, the Boulware vacuum, is not regular on the horizon. 
Warp drive designers might seek to have the
spaceship modulate the quantized field in such a manner
that it would locally, near the horizon, appear to be 
in a state which is regular there. They might eject particles
or otherwise manipulate the field to simulate 
the Hartle-Hawking state at the appropriate temperature or
the Unruh state near the horizon. However,
since the horizon first appears at an infinite distance when
$v = 1$ and subsequently moves inward, it is difficult to see
how the state of the quantized field (presumably of {\it all}
massless fields in nature) could be manipulated at
such great distances from the ship.

Finally, one might object that the divergence perhaps only occurs
along the single spatial direction in which the ship is traveling,
since that is the only direction included in this two-dimensional
calculation. A full four-dimensional calculation would be
needed to settle this issue definitively. 

This research was supported in part by National Science Foundation
Grant No. PHY-9511794. The author wishes to thank Z. Cochrane and S.
Larson for helpful discussions.

\end{document}